\documentclass{sig-alternate}
\usepackage[usenames, dvipsnames]{color} % local add / Sal /
\usepackage{caption}
\usepackage{subcaption}
\usepackage{wrapfig}
\usepackage{url}

\newcommand{\nop}[1]{}

\begin{document}

\title{Random Voting Effects in Social-Digital Spaces:\\ A case study of Reddit Post Submissions}


\numberofauthors{1} %  in this sample file, there are a *total*

\renewcommand{\thefootnote}{\fnsymbol{footnote}}
\author{
% 1st. author
\alignauthor
Maria Glenski$^\dag$ \hspace{.5cm} Thomas J. Johnston$^\ddag$ \hspace{.5cm} Tim Weninger$^\dag$\\
       \affaddr{$^\dag$University of Notre Dame}\\
       \affaddr{$^\ddag$University of Illinois Urbana-Champaign}\\
       \email{mglenski@nd.edu, johnst26@illinois.edu, tweninge@nd.edu}
}
\maketitle

\begin{abstract}
At a time when information seekers first turn to digital sources for news and opinion, it is critical that we understand the role that social media plays in human behavior. This is especially true when information consumers also act as information producers and editors by their online activity. In order to better understand the effects that editorial ratings have on online human behavior, we report the results of a large-scale {\em in-vivo} experiment in social media. We find that small, random rating manipulations on social media submissions created significant changes in downstream ratings resulting in significantly different final outcomes. Positive treatment resulted in a positive effect that increased the final rating by 11.02\% on average. Compared to the control group, positive treatment also increased the probability of reaching a high rating ($\ge$2000) by 24.6\%. Contrary to the results of related work we also find that negative treatment resulted in a negative effect that decreased the final rating by 5.15\% on average.
\end{abstract}

% % A category with the (minimum) three required fields
% \category{Path Length}{Single Source Shortest Paths}{Human Paths}{Hierarchy}{Category}{Heterogenous Information Networks}{Network Analysis} 
% %A category including the fourth, optional field follows...
 % NOT required for Proceedings

\section{Introduction}

What is becoming known as \textbf{collective intelligence} bares the potential to enhance human potential and accomplish what is impossible individually. Indeed, the collective judgements of social groups have been shown to be remarkably accurate when their averaged judgements are compared with the judgements of an individual. For example, more than a century ago the experiments of Francis Galton determined that the median estimate of a group can be more accurate than estimates of experts\cite{Galton1907}. Surowiecki's book \textit{The Wisdom of the Crowds} finds similar examples in stock markets, political elections, quiz shows and a variety of other fields where large groups of people behave intelligently and perform better than an elite few\cite{Surowiecki2005}. However, other experiments have shown that when individuals' perceptions of quality and value follow the behavior of a group, the resulting \textbf{herd mentality} can be suboptimal for both the individual and the group~\cite{Hirshleifer1995, Lorenz2011}.

In modern, digital society, people frequently rely on the anonymous, aggregate ratings of others to make important decisions. The sheer volume of new information being produced and consumed only increases the reliance that individuals place on anonymous others to curate and sort massive amounts of information. But by relying on the judgements of others, we may be susceptible to malicious ratings with some ulterior motive.

With this vulnerability in mind, recent studies have attempted to determine if and how past ratings affect future ratings and the general opinion of the public. Unfortunately, causal determinations are difficult to assess. In a closely related experiment, Wu and Huberman measured rating behavior in two different online platforms. The first allowed users to see prior ratings before they voted and the other platform hid the prior ratings until after the user voted. They found that when no information about previous ratings or page views are available, the ratings and user-opinions expressed tend to follow regular patterns. However, in cases where the previous ratings were made known, the user-opinions tended to be either neutral or form a polarized consensus. In the latter case, new opinions tend to reinforce previous opinions and thus become more extreme~\cite{Wu2008}. 

In a separate line of work, Sorenson used mistaken omissions of books from the \textit{NY Times} bestsellers list to identify the boost in sales that accompany the perceived popularity of a book's appearance on the list\cite{Sorensen2007}. Similarly, when the download counters for different software labels were randomly increased, Hanson and Putler found that users are significantly more likely to download software that had the largest counter increase\cite{Hanson1996}. Salganik and Watts performed a study to determine the extent to which perception of quality becomes a ``self-fulfilling prophecy.'' In their experiment they inverted the true popularity of songs in an online music marketplace, and found that the perceived-but-false-popularity became real over time\cite{Salganik2008}.

The recent popularity of social networks has led to the study of socio-digital influence and popularity cascades where models can be developed based on the adoption rate of friends ({\em e.g.}, share, retweet). Bakshy {\em et al.}, find that the friendship plays a significant role in the sharing of content\cite{Bakshy2009}. Similarly, Leskovec {\em et al.} were able to formulate a generative model that predicts the size and shape of information cascades in online social networks~\cite{Leskovec2006}. 

Like social networks, online {\em social news} platforms allow individuals to contribute to the wisdom of the crowd in new ways. These platforms are typically Web sites that contain very simple mechanics. In general, there are 4 operations that are shared among social news sites: 

\begin{enumerate}
	\item individuals generate content or submit links to content,
	\item submissions are rated and ranked according to their rating scores,
	\item individuals can comment on the submitted content,
	\item comments are rated and ranked according to their rating scores.
\end{enumerate}

Simply put, social news platforms allow individuals to submit content and vote on the content they like or dislike.

The voting mechanism found in socio-digital platforms provides a type of Web-democracy that is open to all comers. Given the widespread use and perceived value of these voting systems\cite{Gilbert2013}, it is important to consider whether they can successfully harness the wisdom of the crowd to accurately aggregate individual information.

In our study, we determine what effect, if any, post ranking and vote score has on rating behavior. This is accomplished via an {\em in vivo} experiment on the social media Web site, Reddit, by inserting random votes into the live rating system.

Reddit is a social news Web site where registered users can submit content, such as direct posts or links. Registered users can then up-vote submissions or down-vote submissions to organize the posts and determine the post's position on the site; posts with a high vote score ({\em i.e.}, up-votes -- down-votes) are ranked more highly than posts with a low vote score. Reddit is organized into many thousands of ``subreddits,'' according to topic or area of interest, {\em e.g.}, news, science, compsci, datamining, and theoryofreddit. Posts must be submitted to a subreddit. A user that subscribes to a particular subreddit will see highly ranked posts from that subreddit on their frontpage, which Reddit describes as `the front page of the Internet.' 

It is important to note that, unlike other online social spaces, Reddit is not a social {\em network}. the notion of friendship and friend-links, like on Facebook, is mostly absent on Reddit. Although usernames are associated with posts and comments, the true identity of registered users is generally unknown and in many cases fiercely guarded. 

In fact, we attempted to find friendship by looking at user-pairs that frequently reply to each other in comments; unfortunately, more than 99.9\% of the comments were in reply to a user that they had never previously replied to. Thus, we typically refer to Reddit a social non-network, and the vast amount of previous social {\em network} literature does not apply.

Although this is the first in-vivo Reddit experiment, our work is motivated and informed by multiple overlapping streams of literature and build on substantial prior work from multiple fields such as: herding behavior from theoretical and empirical viewpoints~\cite{Salganik2006,Weninger2013}; social influence~\cite{Bakshy2009}; collective intelligence~\cite{Hirshleifer1995,Anderson2012a}; and online rating systems~\cite{Lu2008}.

A recent study by Muchnik \textit{et al} on a small social news Web site, similar to Reddit, found that a single up-vote/like on an online comment significantly increased the final vote count of the treated comment; interestingly, the same experiment also found that a single negative rating had little effect on the final vote count\cite{Muchnik2013}.

We report the results of a large ($N=93,019$) in-vivo experiment on Reddit that up-voted or down-voted posts at random. Based on these experimental treatments we observe the effects that votes have on the final score of a post as a proxy for observing herding effects in social news. Unlike the experimental study performed by Muchnik \textit{et al.}, and other behavioral studies our experiment: 1) manipulates votes of posts rather than comments, 2) leverages Reddit's dynamic, score-based ranking system rather than a time-only ranking system, 3) does not involve friendship or the use of social networks, and 4) randomly delays the vote treatment rather than always performing the treatment immediately upon creation.

\section{Methods}

During the 6 months between September 1, 2013 and January 31, 2014 a computer program was executed every 2 minutes that collected post data from Reddit through an automated two-step process. First, the most recent post on Reddit was identified and assigned to one of three treatment groups: up-treated, down-treated, or control. Up-treated posts were artificially given an up-vote (a +1 rating) and down-treated posts were given a down-vote (a -1 rating). Up-treatment, down-treatment and the control have an equal likelihood of being selected. Vote treated posts are assigned a random delay ranging from no delay up to an hour delay in intervals of 0, .5, 1, 5, 10, 30 and 60 minutes. Second, each post was re-sampled 4 days later and final vote totals were recorded. 

These treatments created a small, random manipulation signalling positive or negative judgement that is perceived by other voters as having the same expected quality as all other votes thereby enabling estimates of the effects of a single vote while holding all other factors constant. This data collection resulted in 93,019 sampled posts, of which 30,998 were up-treated and 30,796 were down-treated; each treatment type was randomly assigned a delay interval with equal likelihood. Treatments were removed from the vote scores before data analysis was performed, {\em i.e.}, up-treated post-scores were decremented by 1 and down-treated post-scores were incremented by 1.

During the experimental time period, Reddit reported that their up-vote and down-vote totals were ``fuzzed'' as an anti-spam measure; fortunately, they certified that a post's score ({\em i.e.}, up-votes minus down-votes) was always accurate. In July of 2014, after the data gathering phase of this experiment had ended, Reddit removed the vote totals from their Web site and replaced it with a semi-accurate points system.

\section{Results}
We first compared the final vote totals of posts in each treatment group. These findings measure the overall effect that up-treatments and down-treatments have on the overall life of a post.

\begin{figure}
\centering
    \begin{minipage}{.48\textwidth}%
    \begin{picture}(140,160)% width and height of the picture
\put(0,0){\includegraphics[width=.5\textwidth]{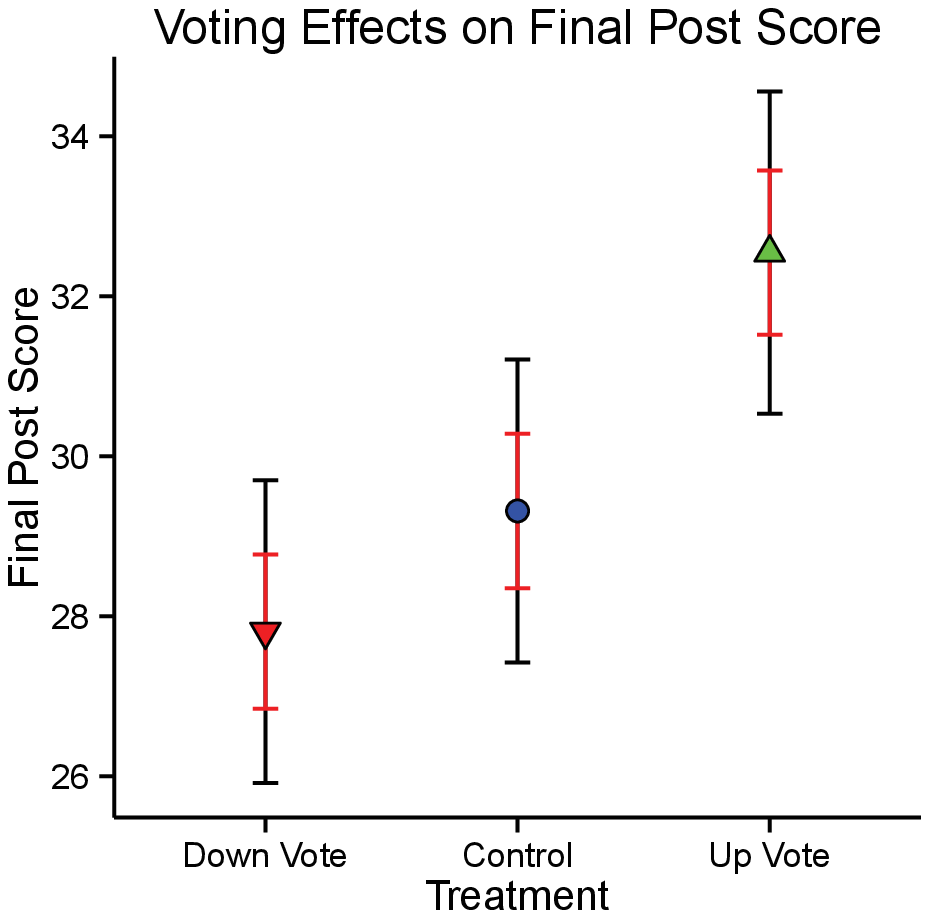}}
\put(100,30){(a)}
\put(120,0){\includegraphics[width=.5\textwidth]{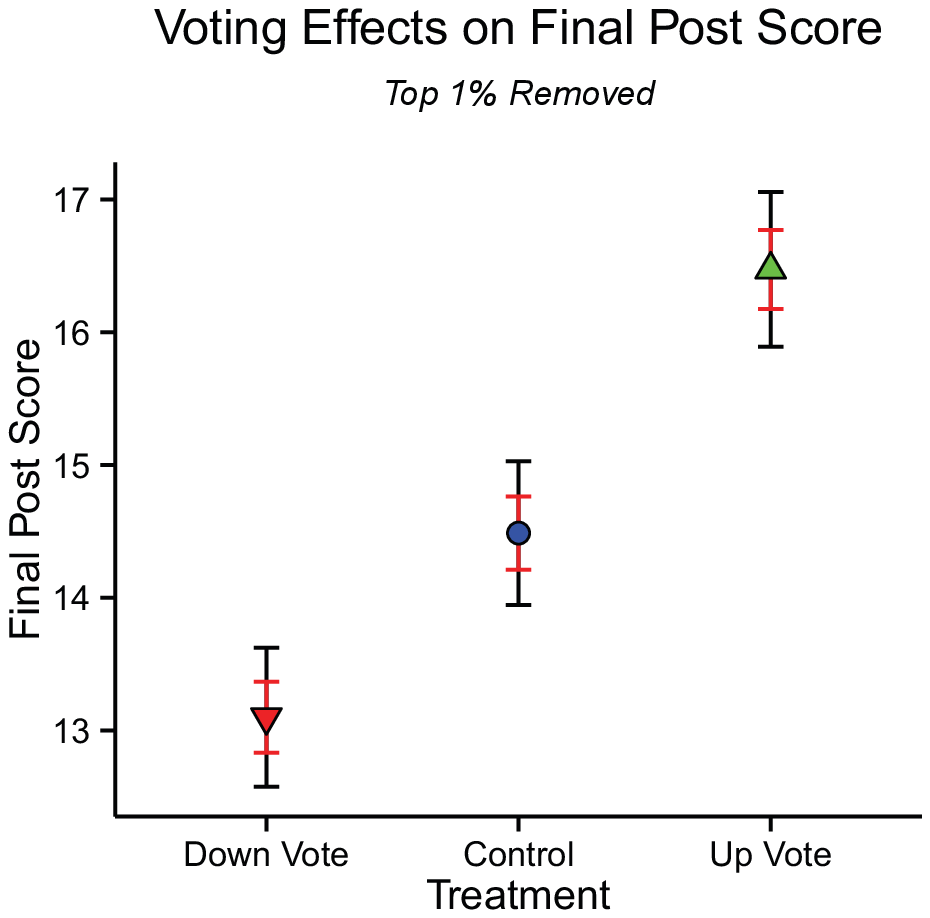}}
\put(220,30){(b)}
\end{picture}
    \end{minipage}%
    
    \caption{Final scores for artificially, randomly up-treated posts, down-treated posts, and scores for untreated posts in the control group are shown. Red inner error bars show the standard error of the mean; black outer error bars show the 95\% confidence interval. Fig. (a) shows the scores in the heavily skewed full distribution. When the highest 1\% of scores are removed as in Fig. (b), the score distribution becomes much less skewed resulting in tighter error bounds, which further result in significant increases for up-treated posts and significant decreases for down-treated posts when compared to the control group.}
    \label{fig:main_res}
\end{figure}

\begin{figure}[t]
\centering
\includegraphics[width=0.48\textwidth]{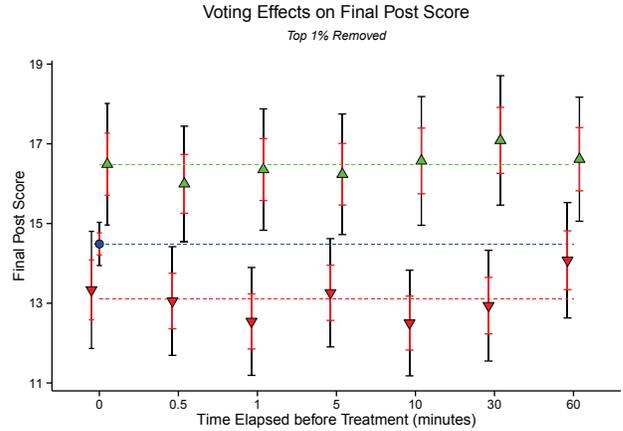}
\caption{Final scores for artificially, randomly up-treated posts, down-treated posts, and scores for untreated posts in the control group separated into their respective treatment delay intervals. Horizontal lines show the overall mean of each treatment group. The top 1\% of scores were removed to un-skew the score distribution. These results show that treatment delay had little effect on the mean final score.}
\label{fig:time_res_99pct}
\end{figure}% 

Figure~\ref{fig:main_res}(a) shows the distribution of the final post scores for each treatment group. Black outer error bars show the 95\% confidence interval and red inner error bars show the standard error of the mean. The distribution of scores is extremely positively skewed with a skewness of $11.2$ and a kurtosis of $149.8$. If we remove the top 1\% highest scoring posts from the data set the skewness and kurtosis values drop to $6.5$ and $54.9$ respectively giving a better, although still skewed, view of the treatment effects. Figure~\ref{fig:main_res}(B) shows the distribution of the final post scores with the top 1\% of posts removed. In this case, the up-treated posts have a significantly higher final score, and the down-treated posts have a significantly lower final score. 

\begin{figure*}[t]
    \centering
    \begin{minipage}[b]{0.19\textwidth}
        \includegraphics[width=\textwidth]{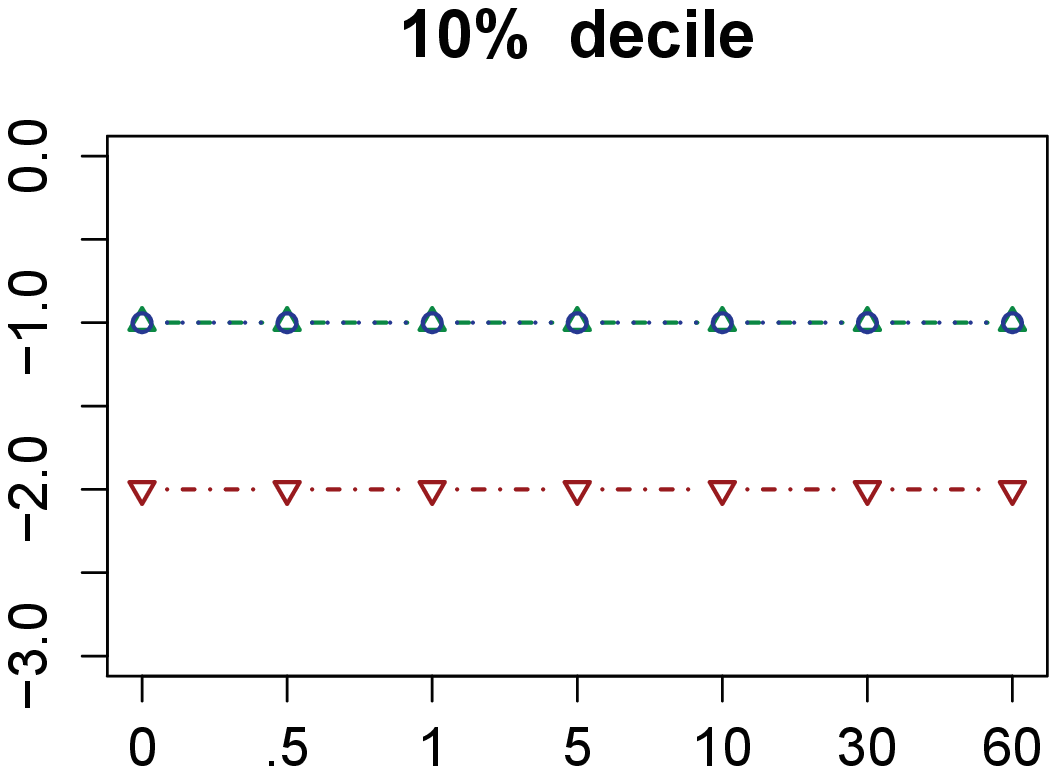}
        %\label{fig:10pctdec}
    \end{minipage}%
    \begin{minipage}[b]{0.19\textwidth}
        \includegraphics[width=\textwidth]{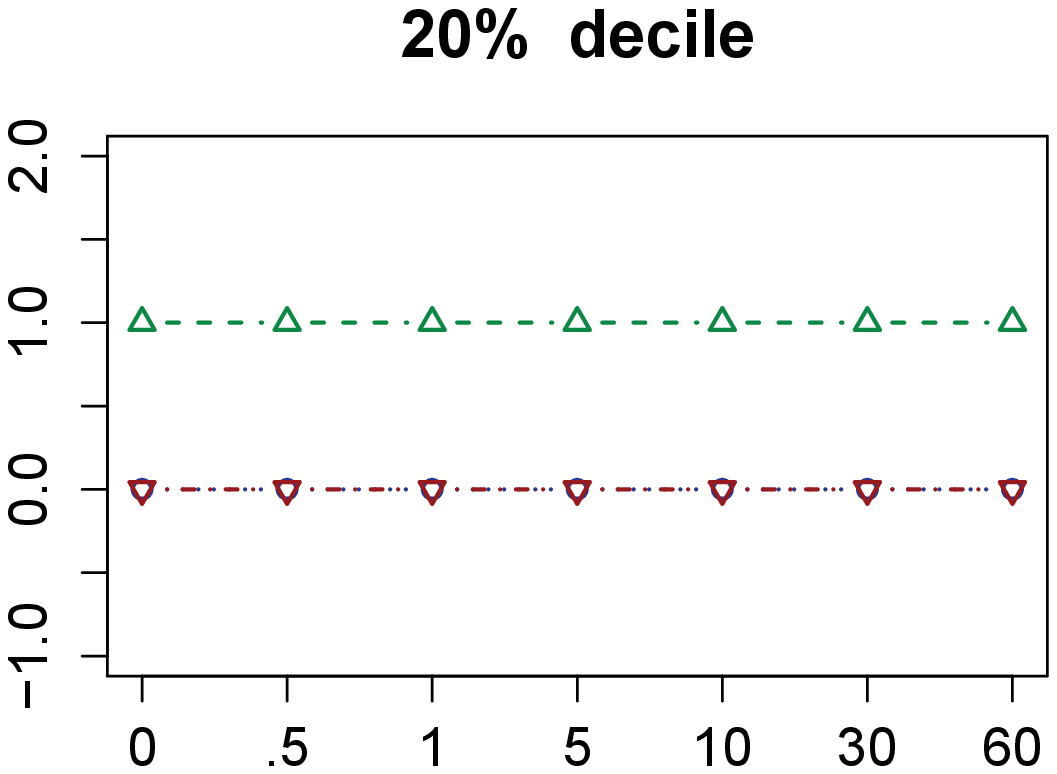}
        %\label{fig:20pctdec}
    \end{minipage}
    \begin{minipage}[b]{0.19\textwidth}
        \includegraphics[width=\textwidth]{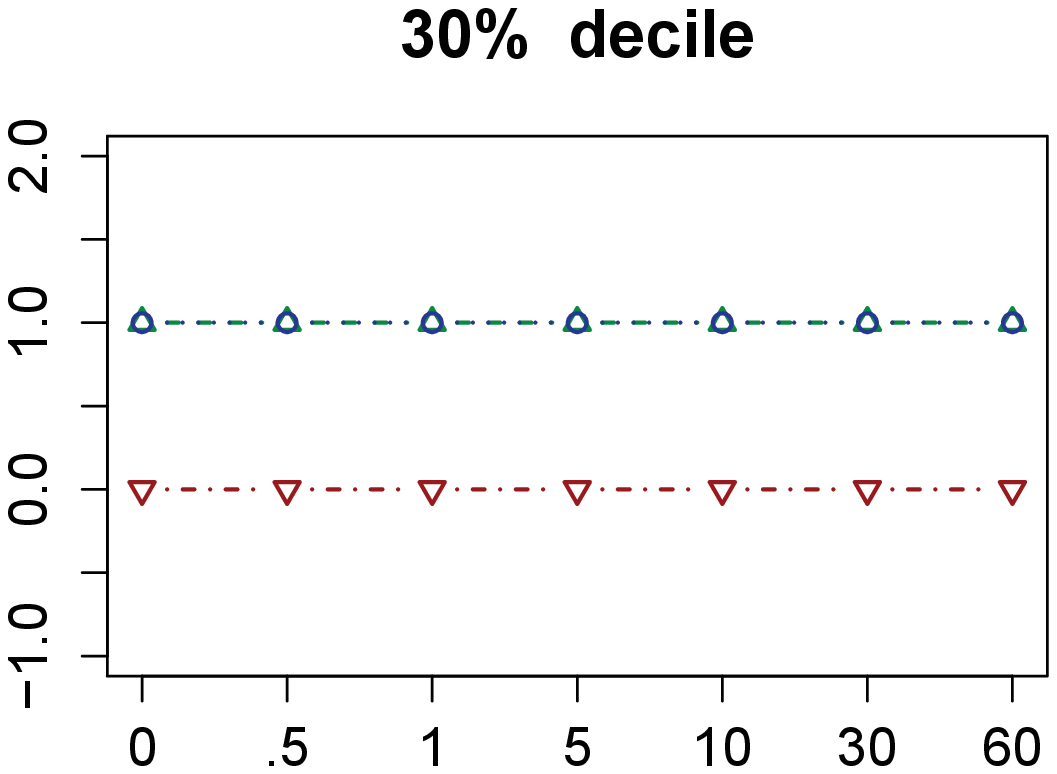}
        %\label{fig:20pctdec}
    \end{minipage}
    \begin{minipage}[b]{0.19\textwidth}
        \includegraphics[width=\textwidth]{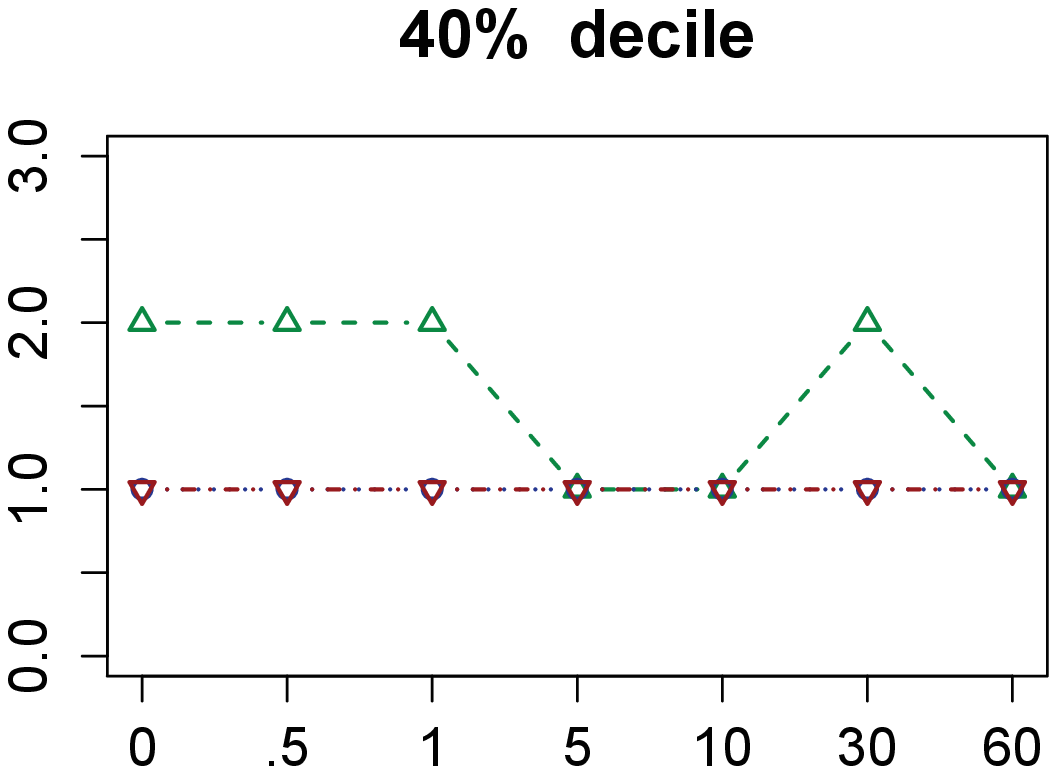}
        %\label{fig:20pctdec}
    \end{minipage}
    \begin{minipage}[b]{0.19\textwidth}
        \includegraphics[width=\textwidth]{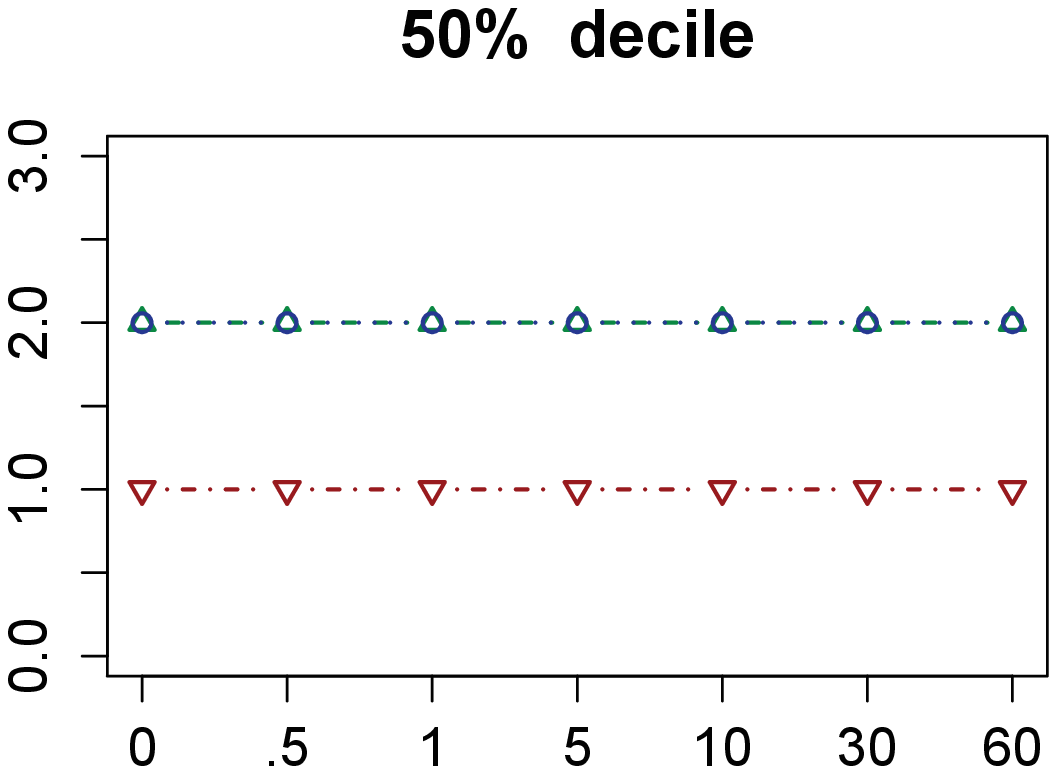}
        %\label{fig:20pctdec}
    \end{minipage}
    \begin{minipage}[b]{0.22\textwidth}
        \includegraphics[width=\textwidth]{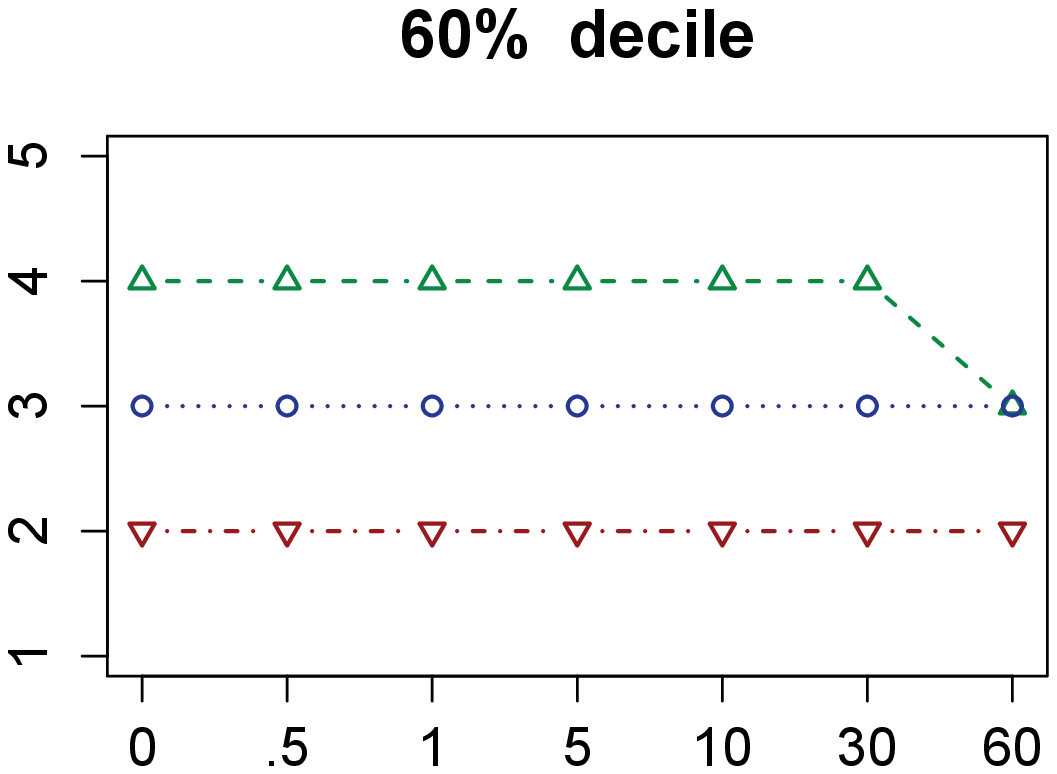}
        %\label{fig:20pctdec}
    \end{minipage}
    \begin{minipage}[b]{0.22\textwidth}
        \includegraphics[width=\textwidth]{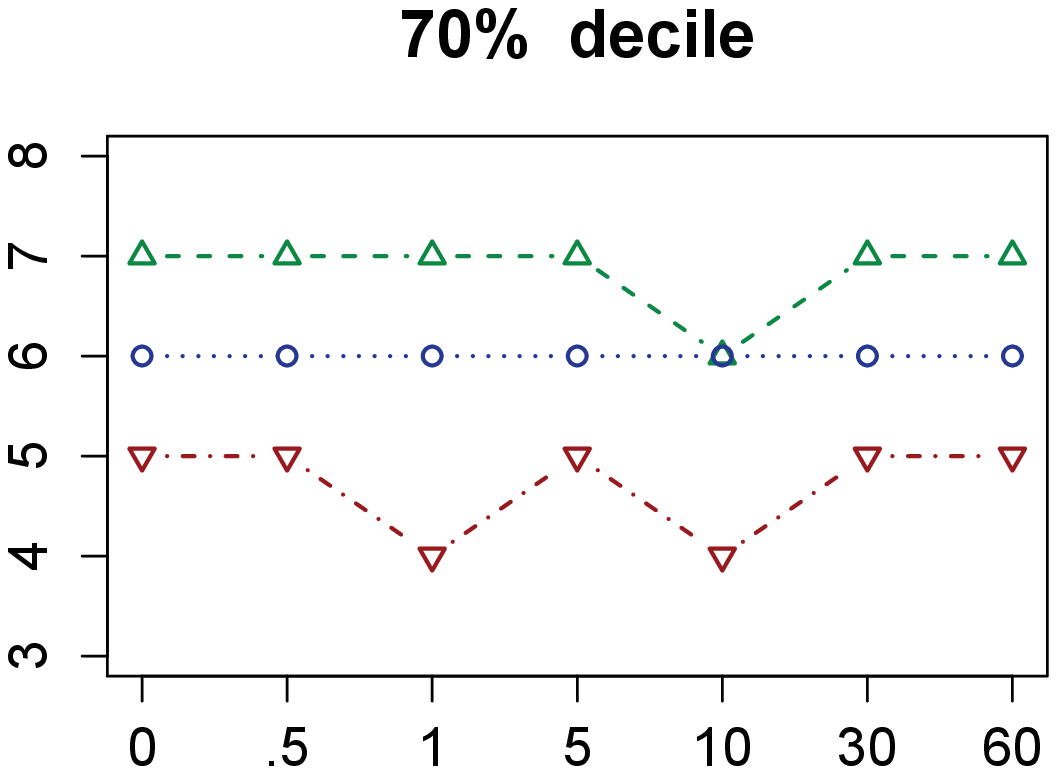}
        %\label{fig:20pctdec}
    \end{minipage}
    \begin{minipage}[b]{0.22\textwidth}
        \includegraphics[width=\textwidth]{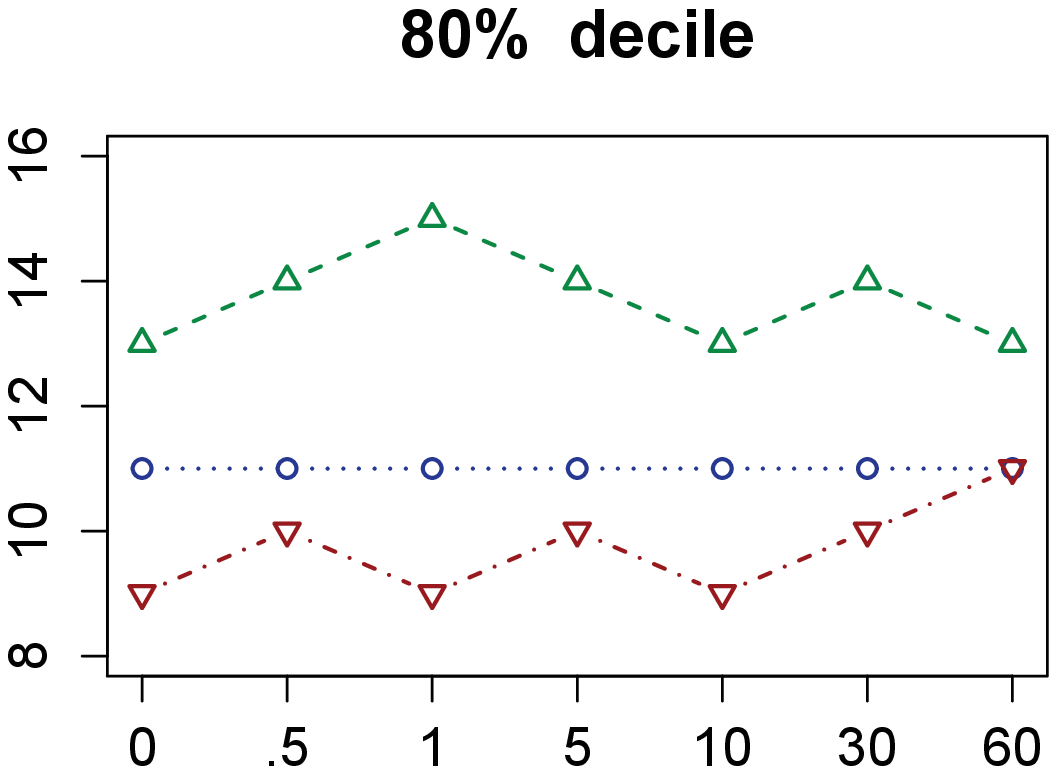}
        %\label{fig:20pctdec}
    \end{minipage}
    \begin{minipage}[b]{0.22\textwidth}
        \includegraphics[width=\textwidth]{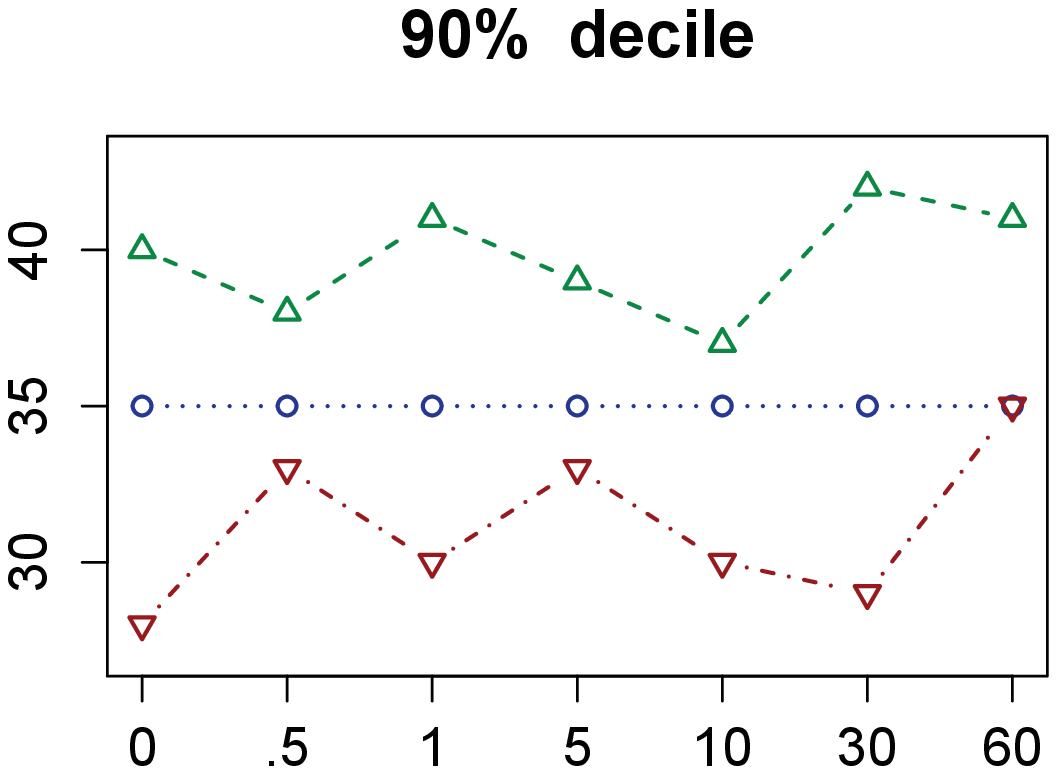}
        %\label{fig:20pctdec}
    \end{minipage}
    \caption{The middle 9 deciles of up-treated, down-treated and control group posts are shown according to their interval times. These results show that most posts receive a median score of 2 or less, and that the treatment has the most effect in the higher deciles of the score distribution.}
    \label{fig:bytime}
\end{figure*}

Kolmogorov-Smirnov (K-S) tests showed that the final score distribution of all up-treated posts were more positively skewed than posts in the control group (K-S test statistic: $0.08$; $p < 2.2 \times 10^{-16}$), which were more positively skewed than down-treated posts (K-S test statistic: $0.11$; $p < 2.2 \times 10^{-16}$). Student's T-Test on log-scores also showed that the up-treated posts were significantly higher than the control group ($p = 1.69 \times 10^{-20}$), and that the down down-treated posts were significantly lower than the control group ($p = 1.69 \times 10^{-09}$) although scores less than or equal to 0 were removed to calculate the log of the final scores.

Up vote and down vote treatments were performed after a 0, 0.5, 1, 5, 10, 30 or 60 minute delay chosen at random, and Figure~\ref{fig:main_res} does not distinguish between the effects of vote-treatments performed after the various delay periods. Figure~\ref{fig:time_res_99pct} separates the results from Figure~\ref{fig:main_res} into their respective treatment delay groups. We expected that immediate votes would have a larger effect than votes performed after a long delay. However, these results show, surprisingly, that a delay in treatment generally had little effect on the mean outcome. Unfortunately, displayed error bounds have little meaning when the data is so highly skewed; K-S tests again showed that all up-treated posts were more positively skewed than posts in the control group and that the effects generally diminished as the delay interval increased: (K-S test statistic: $0.087$(0 min), $0.087$(.5 min), $0.087$(1 min), $0.083$(5 min), $0.082$(10 min), $0.087$(30 min), $0.078$(60 min); $p < 2.2 \times 10^{-16}$).-Similarly, the control group was more positively skewed than the down-treated posts, but the effects were mixed as the delay interval increased: (K-S test statistic: $0.119$(0 min), $0.110$(.5 min), $0.110$(1 min), $0.112$(5 min), $0.119$(10 min), $0.097$(30 min), $0.099$(60 min); $p < 2.2 \times 10^{-16}$). 

Unlike the results presented by Muchnik {\em et al}.~\cite{Muchnik2013} on a similar experiment, we find that the positive and negative treatments show mostly symmetric results, {\em i.e.}, the up-treatments and down-treatments result in similar, yet opposite, departures from the control group. The reason for the differences is unclear and worthy of further investigation.

We are careful to note that reports of mean-averages and standard error are often misleading on such highly skewed data. With this in mind, Figure~\ref{fig:bytime} shows the inner-deciles of the results as a function of their treatment delay. Taken together these results show graphically what the KS tests imply: that up-treated posts tend to be skewed more highly than the control group, and that down-treated posts tend not to be as highly skewed as the control group. The decile plots also show that the majority of posts (deciles $\le$ 50\%) receive at most a final score of 2. A single up-treatment actually does not change the median (50\% decile) final score, but a down-treatment does lower the median score from 2 to 1.

\begin{figure}[t]
\centering
\includegraphics[width=0.48\textwidth]{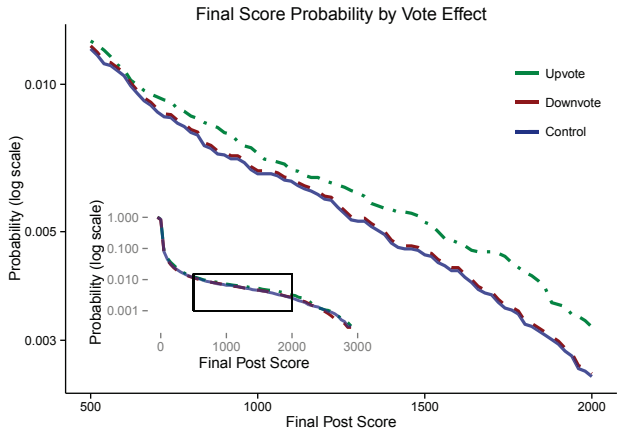}
\caption{ The probability of a post receiving a corresponding score by treatment type. The inset graph shows the complete probability distribution function. The outer graph shows the probability of a post receiving scores between 500 and 2000 -- an approximation for {\em trending} or {\em frontpage} posts. Up-treated posts are 24\% more likely to reach 2000 votes than the control group. }
\label{fig:perc_diff_mean}
\end{figure}

Overall, the results suggest that an up-treatment increases the probability that a post will result in a high score relative to the control group, and that down-treatments decrease that probability relative to the control group. However, on Reddit and other social news sites only a handful of posts become extremely popular. On Twitter and Facebook this is generally referred to as a {\em trending} topic, but on Reddit the most popular posts are the ones that reach the front page. Unfortunately, reaching the front page is a difficult thing to discern because each user's homepage is different, based on the topical subreddits to which the user has subscribed. Nevertheless, we crudely define a post as having become popular, {\em i.e.}, is trending, on the frontpage, etc., if it has a score of more than 500. Using this definition Figure~\ref{fig:perc_diff_mean} shows the probability of a post reaching a given final score under the two treatment conditions. This probability distribution function is monotonically decreasing, positively skewed, and shows that up-treatments result in a large departure from the control group. However, despite our earlier claims of up-treatment and down-treatment symmetry, these results show that, in the upper limits of the distribution, down-treatments do not effect the final score results. These results mean that, compared to the control group, an up-treated post is 7.9\% more likely to have a final score of at least 1000, and an up-treated post is 24.6\% more likely to have a final score of at least 2000.

\begin{figure}[t]
\centering
\includegraphics[width=0.48\textwidth]{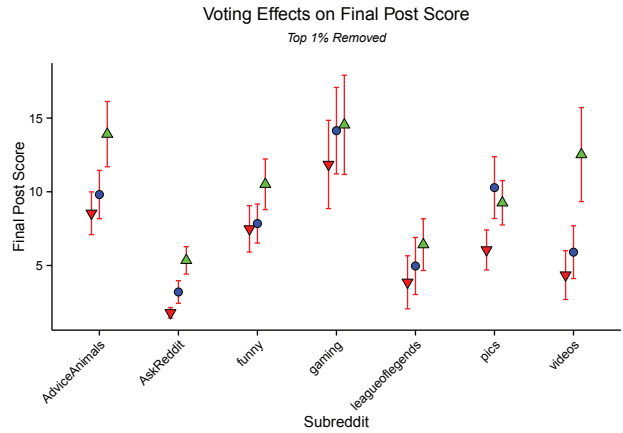}
\caption{ Mean scores of down-treated, control group and up-treated posts shown with 95\% confidence intervals on the top 8 most active subreddits. }
\label{fig:top7}
\end{figure}

We finally investigated treatment effects in the top 10 most frequent subreddits. These do not necessarily correspond to the top 10 most popular subreddits, but are rather the subreddits to which posts are most frequently submitted. From the top 10, we removed \textsf{politic} and \textsf{friendsafari} because posts in \textsf{politic} are automatically submitted by a computer program, and because posts in \textsf{friendsafari} cannot be down-treated according to the subreddit rules. Results show significant positive effects in \textsf{AdviceAnimals}, \textsf{AskReddit} and \textsf{videos}, and significant negative effects in \textsf{AskReddit} and \textsf{pics}. Within the top 500 subreddits, we found that 22\% had significant up-treatment effects, 21.6\% had significant down-treatments, and 5.4\% of subreddits had significant differences in both the up-treatment and down-treatment results as compared to the control group. Also in the top 500 there was no correlation between up-treatment significance and number of submissions the subreddit received ($r^2 = 0.014$; p-value = $0.007$), nor down-treatment significance and number of submissions the subreddit received ($r^2 = 0.010$; p-value = $0.026$)

\section{Discussion}

We find that the positive treatment of a single, random ``upvote'' on a Reddit post has a corresponding positive herding effect that increases post scores on average and in the top limits of the heavily skewed score distribution. We further found that the negative treatment of a single, random ``downvote'' on a post has a corresponding negative herding effect that significantly decreased the post scores on average, in contrast to the asymmetric findings of Muchnik \textit{et al.}~\cite{Muchnik2013}, who found no significant effects of a negative treatment. However, our results begin to resemble asymmetry in the top limits of the score distribution meaning that a negative treatment does not decrease the probability that a post will receive a high score. 

When treatments were separated according to their delay intervals we found that immediate treatments have a slightly larger positive effect than those with longer delays, but the negative delay results did not follow a clear trend. The time of day a vote is placed did not change the overall effect. 

The nature of the manner in which social platforms rank items for viewing typically utilizes the ratings, in this case the post scores, of the items being ranked. The results of our experiment show that random vote perturbations through vote treatments impact the post scores of postings on Reddit. These results underscore the need for counter measures against vote chaining and social engineering strategies as multiple artificial votes are likely to increase the herding effect. 

Finally, we re-draw attention to what Eric Gilbert calls,  the `widespread underprovision of votes' in social media like Reddit~\cite{Gilbert2013}. Although our data does not draw these figures explicitly, we estimate that only .25\% of the of the daily visitors to Reddit actually vote on the items they view. This seems to be an even further skewed anecdote of the 1-9-90 rule of social networking~\cite{vanMierlo2014}, and may be the an underestimated reason behind the results presented in this paper.

Similar work on post comment threads has been collected and will be presented in future reports.

%ACKNOWLEDGMENTS are optional
\section{Acknowledgments}
We thank Michael Creehan for his help and discussion. This research is sponsored by the Air Force Office of Scientific Research FA9550-15-1-0003. The research was approved by University of Notre Dame institution review board and the Air Force Surgeon General's Research Compliance Office. Raw data files, and statistical analysis scripts are available on the corresponding authors Web site at \url{http://www3.nd.edu/~tweninge/data/reddit_report.html}. Reddit Inc was not involved in the experimental design, implementation or data analysis.

\bibliographystyle{abbrv}
%\bibliography{references}  % sigproc.bib is the name of the Bibliography in this case

\balancecolumns
\end{document}